\documentclass[twocolumn,aps,prl]{revtex4}
\usepackage[english]{babel}
\usepackage[dvipdf]{graphicx}
\usepackage[dvips]{epsfig}
\usepackage{amssymb}
\usepackage{amsmath}
\usepackage{amsbsy}
\usepackage{dcolumn}% Align table columns on decimal point

\usepackage[section]{placeins}
\usepackage{color}
\begin{document}

\title{Bursting of dilute emulsion-based liquid sheets driven by a Marangoni effect}

\author{Clara Vernay}
\author{Laurence Ramos}
\email{laurence.ramos@umontpellier.fr}
\author{Christian Ligoure}
\email{christian.ligoure@umontpellier.fr}
%\thanks{ }

\affiliation{
Laboratoire Charles Coulomb UMR 5221,\\
 CNRS, Universit\'{e} de Montpellier,
 F-34095, Montpellier, France\\
}

\date{\today}

\begin{abstract}
We study the destabilization mechanism of thin liquid sheets expanding in air and show that dilute oil-in-water emulsion-based sheets disintegrate through the nucleation and growth of holes that perforate the sheet. The velocity and thickness fields of the sheet outside the holes are not perturbed by holes and hole opening follows a Taylor-Culick law. We find that a pre-hole, which widens and thins out the sheet with time, systematically precedes the hole nucleation. The growth dynamics of the pre-hole follows the law theoretically predicted for a liquid spreading on another liquid of higher surface tension due to Marangoni stresses. Classical Marangoni spreading experiments quantitatively corroborate our findings.

\end{abstract}

\maketitle

The destabilization of free liquid films is of great importance for aerosol dispersions, and is involved in many practical situations~\cite{Herman1987,Bird2010} ranging from foam engineering, food processing, and environmental science. Among the destabilization processes, the disintegration of a liquid sheet through the formation of holes that perforate the liquid film was first mentioned by Dombrowski et al. in the $50$'s ~\cite{Dombrowski1954}, and have been later reported to occur in different types of complex fluids, including surfactant solutions~\cite{Rozhkov2010}, solid suspensions~\cite{Addo-Yobo2011}, surfactant-stabilized air bubbles~\cite{Lhuissier2013}, and dilute oil-in-water emulsions~\cite{Hilz2013}. However, despite its relative ubiquity, the bursting of liquid sheets through perforation events have not yet been carefully investigated nor modeled.
The occurrence of perforation events in a spray directly decreases the fraction of small drops issued from the spray \cite{Vernay2015a} as  illustrated in the case of dilute emulsions,  which are prone to induce the bursting of liquid sheets~\cite{Hilz2013}. Dilute emulsions are also recognized as anti-drift adjuvants  \cite{Ellis1999, Bergeron2003, Qin2010, Vernay2015a} and therefore commonly used as carrier fluids for pesticides delivery. Controlling  the perforation processes would therefore allow one to finely tune the size distribution of drops issued from sprays, a goal of uttermost importance in many industrial processes. In this optics, a physical description of the mechanisms at play in perforation processes is desired. The commonly invoked conditions for perforation include a dewetting of inclusions by the fluid and an inclusion diameter equal to, or larger than, the thickness of the liquid sheet, so that inclusions cause perforation by puncturing both interfaces of the sheet~\cite{Dombrowski1954}. But, to the best of our knowledge, those conditions have never been confronted to robust experimental facts. To unambiguously clarify the physical mechanisms at play, rationale experiments on individual perforation events are therefore required.

In this Letter, we investigate the perforation mechanisms of an emulsion-based free liquid sheet issued from a single-drop experiment; resulting from the impact of one drop of fluid onto a small target \cite{Rozhkov2004, Villermaux2011, Vernay2015}. During the sheet expansion, holes nucleate and grow. We show that each perforation event is preceded by the formation of a pre-hole that thins out the sheet and widens with time. We demonstrate that the pre-hole growth is governed by a Marangoni effect. The entry of emulsion oil droplets at the air/water interface leads to a spreading of the oil due to a surface tension gradient stress. This stress is counterbalanced by a viscous stress that drags the subsurface fluid, whose flow causes a local film thinning which ultimately lead to the rupture of the film.  We show that the growth kinetics of  pre-holes and holes differ. The opening dynamics of holes obeys the Taylor-Culick law \cite{Taylor1959, Culick1960}, whereas the pre-hole dynamics follows the power law evolution predicted for a Marangoni mechanism. Our physical picture is quantitatively confirmed thanks to classical Marangoni experiments where the spreading dynamics of a drop of the emulsion oil phase deposited on a pool of the emulsion aqueous phase is found comparable to that of pre-holes. Thus our paper sheds light on the mechanisms at play during the perforation-driven destabilization of a liquid sheet. A similar Marangoni-driven  mechanism has also been proposed as a possible mode of action of oil-based antifoams \cite{Bergeron1997, Denkov2004}, but to our knowledge has never been experimentally evidenced.

We used dilute oil-in-water emulsions. The oil phase comprises methyllaurate (Sigma-Aldrich) and a mixture of surfactants (provided by Solvay) of type C$_n$EO$_3$, with C$_n$  a hydrocarbon chain with $n$ carbon atoms and EO$_3$  a chain of three ethylene oxide units (CH$_2$CH$_2$O). The aqueous phase is composed of milliQ water (shear viscosity $\eta= 1.0$ mPa.s) or of a mixture of glycerol and water ($62$\%w/w glycerol, $\eta= 12.5$ mPa.s and $75$ \% w/w glycerol, $\eta= 36.5$ mPa.s). The ratio between the oil and the surfactants is set at $97/3$ w/w, and the volume fraction of oil is fixed at $0.3$ \% v/v. A water soluble dye, erioglaucine disodium salt (Sigma-Aldrich), is eventually added to the aqueous phase at a concentration of $2.5$ g/L. The dye does not influence the emulsion surface tension and viscosity, and the oil droplets size distribution. Emulsions are prepared by mechanical stirring, yielding a volume median diameter of the oil droplets of $(20 \pm 4) \, \mu$m, as determined by granulometry.  Details of the experimental set-up are given in \cite{Vernay2015}. In brief, free radially expanding liquid sheets are formed by the impact of a small drop of emulsion (diameter $d_0$=$3.7$ mm) on a  target (chemically treated to avoid dewetting)  of diameter $6$ mm with an impact velocity $u_0=4.0$ m/s. Upon impact, the drop flattens into a radially expanding sheet bounded by a thicker circular rim. The sheet formation and destabilization are recorded from the top with a fast camera (Phantom V7.3) run at an acquisition rate of $10$ kHz. The thickness field of the sheet loaded with dye is determined (range of measurable thicknesses $(5-450)$ $\mu$m with an uncertainty of 5 $\mu$m) thanks to a time- and space-resolved measurement of the adsorbance of the sheet \cite{Vernay2015, Lastakowski2014}.

%-----------------------------------  FIG1 -------------------------------------
\begin{figure}
\includegraphics[width=0.5\textwidth]{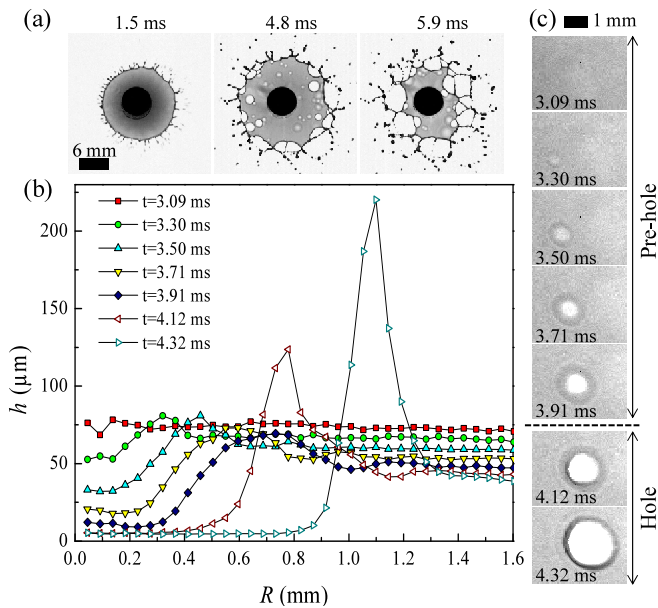}
\caption{(Color online) (a) Formation and destabilization of a dyed emulsion-based liquid sheet. The origin of time, $t$, is taken at the drop impact. (b) Thickness profiles of the patch as a function of the radial distance from the patch center for different times. Open (resp. close) symbols correspond to a pre-hole (resp. hole). (c) Sequence of events of the patch dynamics.}
\label{fig:fig1}
\end{figure}

%------------------------------------ FIG1 --------------------------------------

 Figure~\ref{fig:fig1}(a) displays the destabilization process of a dyed emulsion-based liquid sheet (see movies in the Supplemental Material~\cite{supplementary}). During its expansion, holes perforate the sheet, and grow until they merge forming a web of ligaments, which  are then destabilized into drops. A similar scenario has been observed when a viscous drop impacts a thin layer of liquid with a lower surface tension \cite{Thoroddsen2006a}. For pure water (no oil droplets), no perforation occurs and the sheet uniquely disintegrates through the destabilization of the external rim into  drops~\cite{Rozhkov2004}. Thanks to the dye, the modulations of the sheet thickness are visualized. We find that the formation of a hole is systematically preceded by a localized thinning of the sheet (as depicted by brighter zones in the sheet), referred as pre-hole in the following. Below, we use the generic term "patch" to refer to both a pre-hole and a hole. The frames of fig.~\ref{fig:fig1}(c) illustrate the dynamics of a patch and the corresponding thickness profiles are given in fig.~\ref{fig:fig1}(b). Here the thickness $h$ of the film is plotted as a function of the radial distance $R$ (the origin is taken at the center of the patch) for different times, $t$ (where $t=0$ as the drop impacts the target). Below the numerical values for $t$ and $h$ correspond to the patch considered in fig.~\ref{fig:fig1}(b,c), but the described scenario is robust, and holds for every patch. Typically, about $150$ hole nucleation events occur in one experiment. We have previously shown~\cite{Vernay2015a} that the total number of perforation events is directly correlated to the number of emulsion droplets present in the sheet, proving unambiguously that the emulsions droplets are responsible for the perforation events. At short time, $t=3.09$ ms, the pre-hole is not yet formed and the thickness profile, $h(R)$, is flat with a mean value equal to $71$ $\mu$m. At $t=3.30$ ms, the profile reveals a thinning in the center of the patch, $h(0)=53$ $\mu$m, followed by a smooth bump with a maximal thickness of $81 \, \mu$m corresponding to the rim of the pre-hole. Outside the bump, the thickness profile reaches a plateau, $h_{\rm{out}}$, corresponding to the undisturbed sheet. With time, the pre-hole thins down to $h(0)=12 \,\mu$m at $t=3.91$ ms and widens up to a radius $R_{\rm{ph}}=0.75$ mm. At $t=4.12$ ms, the film ruptures. The pre-hole to hole transition is clearly noticeable by a sharp peak in the thickness profile corresponding to the thicker, i.e. darker, rim surrounding the hole and a vanishing thickness of the inner zone of the patch.

%-----------------------------------  FIG2 -------------------------------------
\begin{figure}
\includegraphics[width=0.5\textwidth]{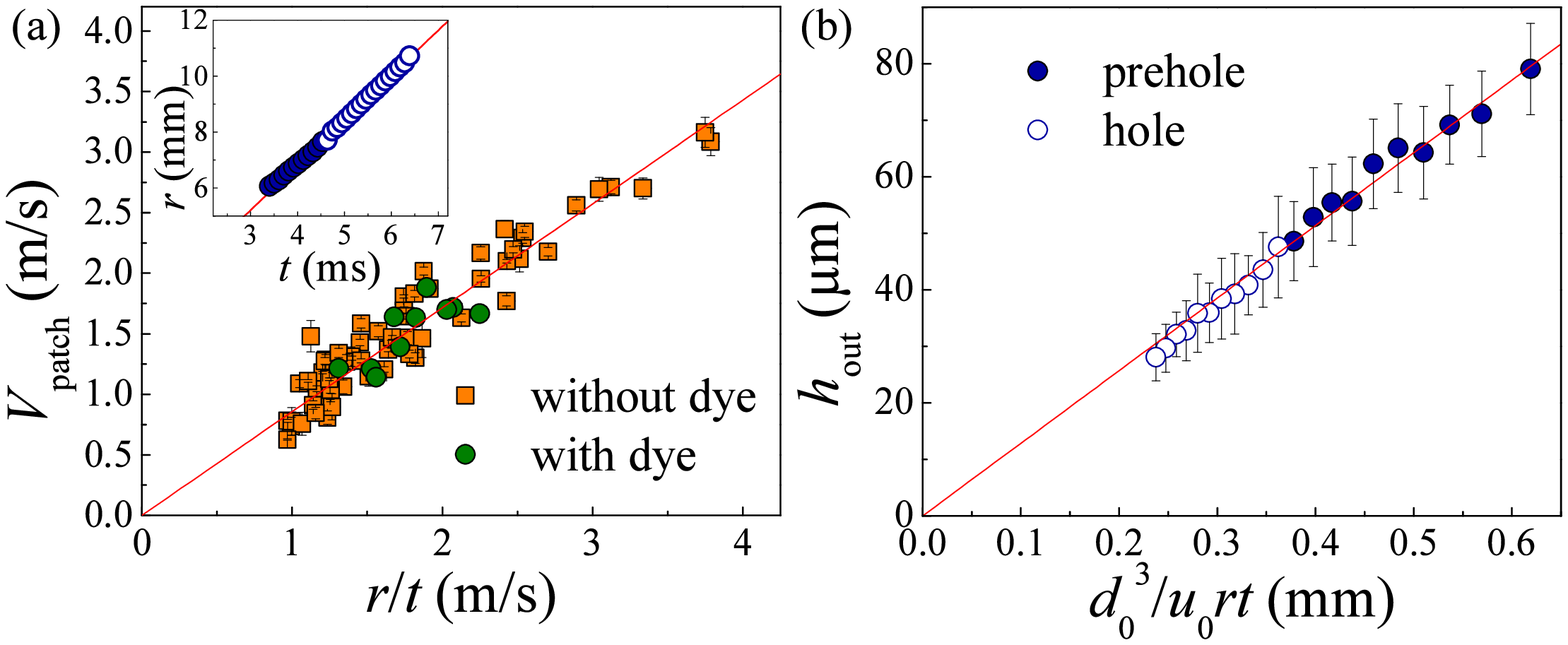}
\caption{(Color online) (inset)  Trajectory of one  patch (filled symbols: pre-hole, open symbols: hole). (a) Evolution of the  radial patch velocity as a function of $r/t$. (b) Evolution of the sheet thickness in the vicinity  of a patch with $\frac{{d_0}^3}{u_0 rt}$, where $r$ is the radial distance from the center of the target, and $t$ the time from impact. The red lines in (a,b) are the best linear fits.}
\label{fig:fig2}
\end{figure}
%------------------------------------ FIG2 --------------------------------------

We first discuss the overall dynamics of the sheet. Inset of fig.~\ref{fig:fig2}(a) displays the time evolution of the position of one patch. We measure that the radial position of each patch, $r$, (the origin of $r$ is the center of the target) varies linearly with time, indicating a constant radial speed within the sheet, $V_{\rm{patch}}$. No discontinuity at the perforation transition is depicted, as data corresponding respectively to pre-hole (filled circles, inset of fig.~\ref{fig:fig2}(a)) and to hole (open circles) follow a unique straight trajectory. Moreover, we find that for all patches, $V_{\rm{patch}}$ is proportional to $r/t$, (fig.~\ref{fig:fig2}(a)). This scaling is the one expected for a plain unperturbed expanding liquid sheet~\cite{Villermaux2011} with an expected prefactor of $1.0$, comparable to the experimental one ($0.86$)~\cite{Rozhkov2004,Villermaux2011}. The quantitative correspondence with unperturbed plain liquid sheets is further confirmed by measurements of the thickness field. Figure~\ref{fig:fig2}(b) shows the evolution of the thickness of the film in the outside vicinity of the patch, $h_{\rm{out}}$, with $\frac{{d_0}^3}{u_0 rt}$. The plot shows the continuity of $h_{\rm{out}}$ at the pre-hole to hole transition, and that all over the process, the thickness is proportional to $\frac{{d_0}^3}{u_0 rt}$, with a proportionality constant $\alpha=(0.094 \pm 0.018)$. This thickness field corresponds quantitatively to that of an expanding  plain water  sheet (without holes) as  measured ~\cite{Vernay2015} and  predicted~\cite{Villermaux2011}. The theoretical proportional constant $\alpha=1/12=0.083$ \cite{Villermaux2011}, is  in very good agreement with the experimental one.  This result indicates that the  perforation of the sheet does not perturb the thickness field of the sheet outside the patches. Therefore each pre-hole can be viewed as a closed system, such that all the liquid evacuated from the pre-hole thinning zone is transferred in its rim. Overall, the results of fig.~\ref{fig:fig2} conclusively demonstrate that the kinematic fields of the sheet outside of the patches are not perturbed neither by the presence of oil droplet in the sheet, nor by the perforation events.

%-----------------------------------  FIG3 -------------------------------------
\begin{figure}
\includegraphics[width=0.5\textwidth]{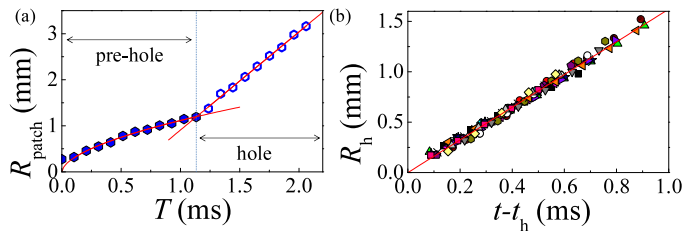}
\caption{(Color online) (a)  Evolution of the radius of the patch (filled symbols: pre-hole, open symbols: hole) with the time elapsed since the formation of the pre-hole, $T$, for the patch shown in fig.~\ref{fig:fig1}(b,c). The solid lines are the best power law fit, resp. linear fit, for the the pre-hole, resp. hole, dynamics. (b) Evolution of the radius of the holes with the time elapsed since the hole formation. Data for more than $20$ holes are plotted, each color corresponding to a given hole. The red line is the best linear fit.}
\label{fig:fig3}

\end{figure}
%------------------------------------ FIG3 --------------------------------------

To elucidate the physical mechanisms at play in the bursting process, we investigate the growth dynamics of the patches. The evolution of the patch radius, $R_{\rm{patch}}$, as measured at the maximum of the thickness profile, with the time elapsed since the pre-hole formation, $T=t-t_{\rm{ph}}$ (here $t_{\rm{ph}}$ corresponds to the time at which the pre-hole forms) is shown in fig.~\ref{fig:fig3}(a).
The pre-hole to hole transition event, i.e. the rupture of the film, is  clearly evidenced by a discontinuity in the dynamics of $R_{\rm{patch}}$. This discontinuity separates two distinct regimes, corresponding to the pre-hole and hole growths. We first focus on the second regime. Figure~\ref{fig:fig3}(b) displays the evolution of the hole radii, $R_{\rm{h}}$, for more than twenty holes with the time elapsed since the nucleation of the hole, at time $t_{\rm{h}}$. We find that $R_{\rm{h}}$ increases linearly with time, with a constant opening velocity $V_{\rm{c}}=(1.64 \pm 0.16)$ m/s, which results from a balance between the rim inertia and surface tension in the film as predicted by Taylor~\cite{Taylor1959} and Culick~\cite{Culick1960} for the rupture of a soap film: $V_{\rm{c}}=\sqrt{2\gamma/(\rho h)}$, with $\rho$ the density of the liquid, $\gamma$ its surface tension, and $h$ the film thickness. Although a constant velocity is not expected here as the thickness decreases with time~\cite{Lastakowski2014}, we quantitatively show in the Supplemental Material~\cite{supplementary} that this effect is negligible for the time- and length-scales considered here. Numerically, one considers for $h$ the thickness experimentally measured in the outside periphery of the hole, $h_{\rm{out}}=\frac{{\alpha d_0}^3}{u_0 rt}$ with $\alpha=(0.094 \pm 0.018)$  (fig.~\ref{fig:fig2}(b)) taking for $r$ and $t$ the values at the hole formation. With $\gamma =70.1$ mN/m, the surface tension of the emulsion measured at short times ($\leq50$ ms), and $\rho = 998$ kg/m$^3$, one predicts $V_{\rm{c}}=(1.68 \pm 0.12)$ m/s, a numerical value in excellent agreement with the experimental value~\cite{note}.

%-----------------------------------  FIG4 -------------------------------------
\begin{figure}
\includegraphics[width=0.5\textwidth]{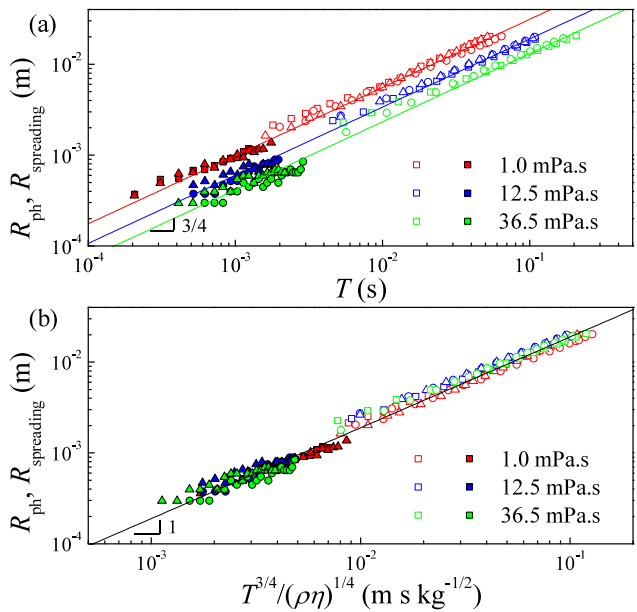}
\caption{(Color online) (a) Evolution of the radius of the pre-hole, $R_{\rm{ph}}$, in the single drop experiment (full symbols), resp. of the spreading radius, $R_{\rm{spreading}}$, in the standard Marangoni experiments (open symbols), as a function of $T$, the time elapsed since the pre-hole formation, resp. since the deposition of an oil drop. Different symbol types correspond to different experiments. The lines are the best fits of the experimental data of the form $R_{\rm{ph}}=kT^{3/4}$. (b) Same data as in (a) plotted as a function of $T^{3/4}/(\rho\eta)^{1/4}$.}
\label{fig:fig4}
\end{figure}
%------------------------------------ FIG4 --------------------------------------

The evolution of the pre-hole radius, $R_{\rm{ph}}$, with the time elapsed since its formation, $T=t-t_{\rm{ph}}$, is plotted in  Fig.~\ref{fig:fig4}(a) (full symbols) for three different aqueous phases, whose zero-shear viscosity varies between $1$ and $36.5$ mPa.s.  In all cases, $R_{\rm{ph}}$ increases with $T$ as a power law with an exponent $3/4$: $R_{\rm{ph}}=kT^{\frac{3}{4}}$. This scaling suggests a widening dynamics in agreement with the dynamics of Marangoni spreading, i.e. the spontaneous spreading of a thin film (of say oil phase) along the surface of a deep fluid layer (of say aqueous phase) of higher surface tension~\cite{Fay1969, Joos1977, Foda1980}. The driving stress of the spreading is the surface tension gradient associated with the presence of an oil drop at the interface between air and the liquid film. This driving stress leads  to the spreading of the oil drop at the interface inducing   a viscous shear stress that causes the liquid in the film to flow in the direction of the surface tension gradient resulting in the deformation of the interface with a localized thinning of the film. We argue that such mechanism can be at the origin of the pre-hole formation and growth.

To confirm this statement, we perform classical Marangoni spreading experiments \cite{Joos1985, Camp1987, Bergeron1996}. Talcum particles are first sprinkled on the surface of an undisturbed pool of aqueous phase (thickness $1$ cm). The subsequent deposition of a droplet (4 $\mu$L) of the oil phase yields a radial spreading of the oil, quantified thanks to the particles motion. The evolution of the radius of the spreading area, $R_{\rm{spreading}}$ is plotted in fig.~\ref{fig:fig4}(a) (open symbols) as a function of time $T$ (the origin is taken at the moment of the deposition of the oil drop at the interface) for the different viscosities of the bulk liquid. Hence for both types of experiments (single drop and classical Marangoni), $T=0$ corresponds to the beginning of the spreading. Similarly to the single drop experiments, the data for $R_{\rm{spreading}}$ follow a power law with an exponent $3/4$. Although the time scales involved in the two types of experiments are significantly different, the data points for $R_{\rm{ph}}$ and $R_{\rm{spreading}}$ lye on a unique curve, suggesting a same prefactor, $k$, and showing that a same mechanism drives the two processes. Our results show moreover that $k$ decreases with viscosity \cite{Comment}.  For surface-tension gradient-driven spreading, the prefactor $k$ is expected to vary as $k=K S^{1/2}/(\rho \eta^{1/4})$, where $K=\sqrt{4/3}$, and $\rho$ and $\eta$ are the density and viscosity of the bulk aqueous liquid~\cite{Dussaud1998a}. $S$ is the spreading coefficient defined as $S=\gamma_{\rm{air/aq}}-\gamma_{\rm{air/oil}}-\gamma_{\rm{aq/oil}}$, where $\gamma_{\rm{a/b}}$ stands for the interfacial tension between phases "a" and "b", and "aq" stands for the aqueous phase. To quantitatively check the scaling of the prefactor $k$, $R_{\rm{ph}}$ and $R_{\rm{spreading}}$ are re-plotted as a function of the rescaled parameter  $T^{3/4}/{\rho\eta}^{1/4}$ in fig.~\ref{fig:fig4}(b). Remarkably, all the data acquired in the two experiments and for different viscosities of the aqueous phases fall on a unique master curve, with a slope of $1$, as expected. A spreading parameter $S$ can thus be readily extracted from the slope of the linear evolution of the spreading radius with $T^{3/4}/{\rho\eta}^{1/4}$ (fig.~\ref{fig:fig4}(b)). One finds $S=(28.2 \pm 4.8)$ mN/m. This value can be directly compared to macroscopic measurements of the spreading parameter. With $\gamma_{\rm{air/aq}} = (67.5 \pm 4.1)$ mN/m, $\gamma_{\rm{air/oil}}= 29.5$ mN/m, and $\gamma_{\rm{aq/oil}}= 1.1$ mN/m, one estimates $S=(36.9 \pm 4.1)$ mN/m. Hence, $S > 0$, a prerequisite for Marangoni spreading. The values derived from direct measurements and the one derived from the master curve are in reasonable agreement, considering in particular the fact that the numerical value of $K$ is subject to discussion. Here we take $K=\sqrt{4/3} \simeq 1.15$ but numerical values between $0.665$ and $1.52$ have been mentioned in the literature~\cite{Dussaud1998a}.

We argue that the formation of  a pre-hole is due to the entering from the solvent sub-phase of emulsion oil droplets that subsequently spread at the air/water interface leading to the local thinning of the sheet, because of a Marangoni surface tension gradient, which eventually leads to the film rupture. The deep pool assumption, for which the power law exponent of the spreading is $3/4$, holds when the penetration length $\delta=\sqrt{\frac{\eta T}{\rho}}$ is smaller than the thickness of the sub-phase $h$. This is always the case for the classical Marangoni spreading experiments. For the single-drop experiment,  this assumption is not fulfilled during the whole duration of the thinning process  for the more viscous samples ($\eta=12.5$ and $36.5$ mPa.s); however we do not measure significative deviation from the spreading law predicted in the deep pool regime, as one would expect once entering the thin-film regime ($\delta \sim h$) \cite{Ahmad1972}. Hence, the rupture of the film seems to occur in the spreading regime predicted by the deep pool assumption. One possible reason is that the liquid sheet expands freely in air (free boundary conditions) contrary to the assumptions in thin film models. We experimentally determine the time to rupture as the time elapsed from the pre-hole formation to the film rupture: $(1.15 \pm 0.44)$ ms, and we find that this time is comparable to $\frac{\rho h_0^2}{\eta}=(2.8  \pm  0.07)$ ms the time at which $\delta\sim h_0$, with $h_0=(52.6 \pm 12.8) \mu$m, the thickness of the sheet when the pre-hole forms. However a rationale theoretical description of the ultimate stage of the rupture  is still lacking. Note finally that pre-hole formation occurs when the thickness of the sheet  is significantly larger that the size of the oil droplets ($\sim 20 \, \mu$m), definitely ruling out a puncture mechanism.

To conclude, we have provided a rational description of the dynamics of pre-holes and holes occurring in free liquid sheets comprising oil droplets and have quantitatively proven that hole nucleation is governed by Marangoni spreading of the oil droplets at the air/water interface.

\begin{acknowledgments}
We thank P. Perrin and M. Protat for their help in the measurement of the dynamic surface tension, and E. Villermaux, J.-C. Castaing and I. Cantat for fruitful discussions. Financial support from Solvay is acknowledged.
\end{acknowledgments}

%--------------------------------------------- REFERENCES -----------------------------------------

%\bibliographystyle{aipnum4-1}
%\bibliography{Article_marangoni}

\begin{thebibliography}{26}
\bibitem{note} Note that the experimental uncertainty on the opening velocity can be fully accounted for by the variation of $h_{\rm{out}}$, the thickness in the outside periphery of the hole, for the holes considered in fig.~\ref{fig:fig3}(b).
\bibitem[by contrast, the opening velocity of the hole is found independent of the viscosity of the subphase as expected from Taylor-Culick law ]{Comment}
  \bibinfo{journal}{By contrast, the opening velocity of the hole is found independent of the viscosity of the subphase, as expected from the Taylor-Culick law.}
\end{thebibliography}

\end{document}